\documentclass[10pt,final,journal,a4paper,oneside,twocolumn]{IEEEtran}
\usepackage{srcltx}
\usepackage{psfrag}
\usepackage{epsfig}
\usepackage{color}
\usepackage{ulem}
\usepackage[tbtags,sumlimits,nointlimits,reqno]{amsmath}
\usepackage{amssymb}
\usepackage{showkeys}   
\usepackage{float}
\usepackage{subfigure}
\usepackage{type1cm}
\usepackage{lettrine}
\usepackage{epstopdf}
\usepackage{pstool}
\usepackage{cite}
\begin{document}
\title{Enhanced-SNR Impulse Radio Transceiver \\ based on Phasers}
%
\author{%
       Babak~Nikfal,~\IEEEmembership{Student Member,~IEEE,} Qingfeng~Zhang,~\IEEEmembership{Member,~IEEE,} and Christophe~Caloz,~\IEEEmembership{Fellow,~IEEE}
}
\markboth{Microwave and Wireless Component Letters~2014}{Shell \MakeLowercase{\textit{et al.}}: Bare Demo of IEEEtran.cls for
Journals}
%
\maketitle
\begin{abstract}
The concept of SNR enhancement in impulse radio transceivers based on phasers of opposite chirping slopes is introduced. It is shown that signal-to-noise radio (SNR) enhancements by factors $M^2$ and $M$ are achieved for burst noise and Gaussian noise, respectively, where $M$ is the stretching factor of the phasers. An experimental demonstration is presented, using stripline cascaded C-section phasers, where SNR enhancements in agreement with theory are obtained. The proposed radio analog signal processing transceiver system is simple, low-cost and frequency scalable, and may therefore be suitable for broadband impulse radio ranging and communication applications.
\end{abstract}
\section{Introduction}
Impulse radio, a form of ultra-wide band signaling using pulse modulation, is a promising technology for high data rate communication networks with low complexity and low power consumption. It may for instance apply to wireless personal area networks, identification, positioning and wireless sensor networks~\cite{Impulse-Radio-1,Impulse-Radio-2,Impulse-Radio-3}. However, impulse radio systems are particularly sensitive to noise and interferers since they are fundamentally amplitude modulation system. Increasing the power of the transmitted signal is not a solution for SNR enhancement, since it leads to increased power consumption, reduced system dynamic range due to the power amplifier, and exceeding of allowed power spectral density (PSD) limits.

To address this issue, we introduce here an enhanced-SNR impulse radio transceiver, based on the up-chirp and down-chirp dispersive delay structures with controllable group delay~\cite{Zhang_APM_05_2013,Nikfal_loop,Zhang_TMTT_03_2013}, called phasers~\cite{Caloz-Metamaterial,ASP-Magazine}. Due to its passive nature and broadband nature of phasers, enhanced-SNR impulse radio transceiver is simple, broadband and frequency scalable.
\section{Principle of the Transceiver}

\begin{figure}[h!]
\begin{center}
\subfigure[]{\label{Fig:tx}
\centering
\psfragfig*[width=0.95\columnwidth]{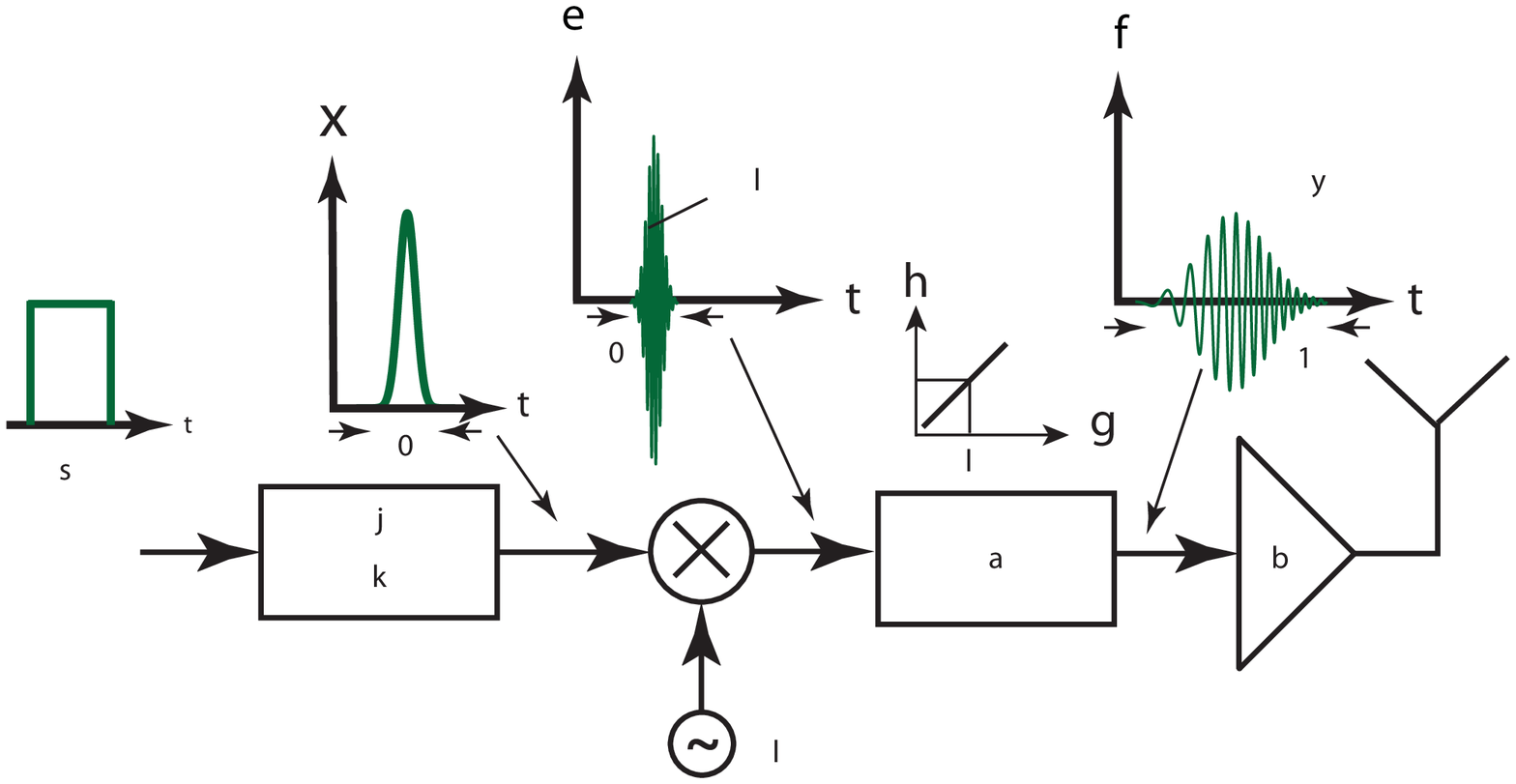}
{\psfrag{h}[c][c][0.7]{$\tau$}
\psfrag{t}[c][c][0.7]{$t$}
\psfrag{j}[c][c][0.7]{pulse}
\psfrag{k}[c][c][0.7]{shaper}
\psfrag{l}[c][c][0.7]{$\omega_0$}
\psfrag{a}[c][c][0.7]{$\tau(\omega)$}
\psfrag{b}[c][c][0.7]{PA}
\psfrag{g}[c][c][0.7]{$\omega$}
\psfrag{x}[c][c][0.7]{$g(t)$}
\psfrag{e}[c][c][0.7]{$v^-_\text{Tx}(t)$}
\psfrag{f}[c][c][0.7]{$v^+_\text{Tx}(t)$}
\psfrag{y}[c][c][0.7]{up-chirped signal}
\psfrag{0}[c][c][0.7]{$T_0$}
\psfrag{1}[c][c][0.7]{$MT_0$}
\psfrag{s}[c][c][0.7]{input pulse}}}
\subfigure[]{\label{Fig:rx}
\centering
\psfragfig*[width=0.95\columnwidth]{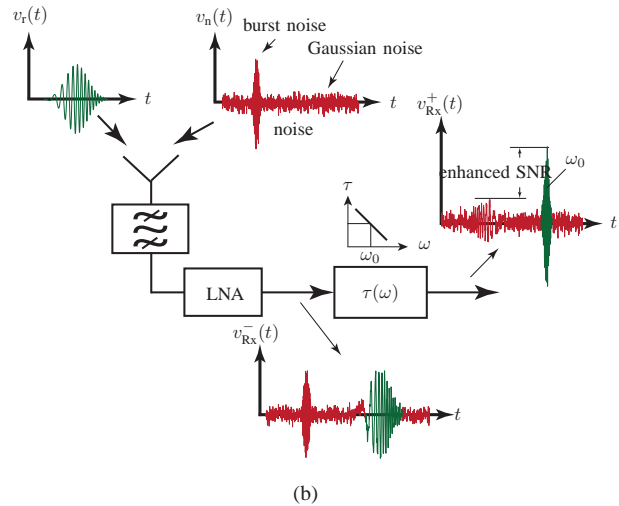}
{\psfrag{t}[c][c][0.7]{$t$}
\psfrag{h}[c][c][0.7]{$\tau$}
\psfrag{g}[c][c][0.7]{$\omega$}
\psfrag{n}[c][c][0.7]{$v_\text{n}(t)$}
\psfrag{m}[c][c][0.7]{Gaussian noise}
\psfrag{l}[c][c][0.7]{burst noise}
\psfrag{i}[c][c][0.7]{$\omega_0$}
\psfrag{a}[c][c][0.7]{$\tau(\omega)$}
\psfrag{p}[c][c][0.7]{LNA}
\psfrag{o}[c][c][0.7]{$v^+_\text{Rx}(t)$}
\psfrag{y}[c][c][0.7]{$v^-_\text{Rx}(t)$}
\psfrag{x}[c][c][0.7]{noise}
\psfrag{f}[c][c][0.7]{$v_\text{r}(t)$}
\psfrag{r}[c][c][0.7]{enhanced SNR}}}
\caption{Principle of the proposed enhanced-SNR impulse radio transceiver. (a)~Transmitter. (b)~Receiver.}
\label{Fig:block}
\end{center}
\end{figure}

Figure~\ref{Fig:block} shows the principle of the proposed enhanced-SNR impulse radio transceiver. Figure~\ref{Fig:tx} shows the block diagram of the transmitter. The message data, which are reduced in the figure to a single baseband rectangular pulse representing a bit of information, is injected into the pulse shaper to be transformed into a smooth Gaussian-type pulse, $g(t)$, of duration $T_0$. This pulse is mixed with an LO signal of frequency $\omega_0$, which yields the modulated pulse $v^-_\text{Tx}(t)=g(t)\cos(\omega_0t)$, of peak power $P_0$. This modulated pulse is injected into a linear up-chirp phaser, and subsequently transforms into an up-chirped pulse, $v^+_\text{Tx}(t)$. Assuming energy conservation (lossless system), the duration of this pulse has increased to $MT_0$ while its peak power has decreased to $P_0/M$, where $M$ is the stretching factor of the phaser~\cite{nikfal-frequency-sniffer}\footnote{$M=T_\text{out}/T_0=(T_0+s\Delta\omega)/T_0$, where $T_0$ and $T_\text{out}$ are the duration of the input and output pulse, repectively, $s$ is the slope of the group delay response of the phaser (in s$^2/$rad), $\tau(\omega)$, and $\Delta\omega$ is the bandwidth of the pulse~\cite{nikfal-frequency-sniffer}.}. Finally, the up-chirped pulse $v^+_\text{Tx}(t)$ is boosted by a power amplifier and radiated by an antenna towards the receiver.

Figure~\ref{Fig:rx} shows the block diagram of the receiver. The incoming signal, $v_r(t)$, is picked up by the antenna, along with the noise of the channel, $v_n(t)$, which may consist of a superposition of Gaussian noise and burst noise. After bandpass filtering and low-noise amplification, the received up-chirped and noisy signal, $v^-_{Rx}(t)$, is injected into a linear down-chirp phaser~\footnote{Alternatively, one may naturally use a down-chirp phaser in the transmitter and an up-chirp signal in the receiver.}. As a result, the instantaneous frequencies of the transmitted part of the signal are essentially equalized so as to compress the pulse to an enhanced waveform that is essentially identical to $v^-_{Tx}(t)$, whereas the burst noise is spread out in time, since it was not pre-chirped, and the Gaussian noise remains a Gaussian noise. So, the signal has been enhanced whereas the noise level has been either reduced (burst noise) or unaffected (Gaussian noise). Note that time expansion and associated peak power reduction in the transmitter also results in increased system dynamic range.

\section{Signal-to-Noise Characterization}

\begin{figure}[h!]
\centering
\psfragfig*[width=.85\columnwidth]{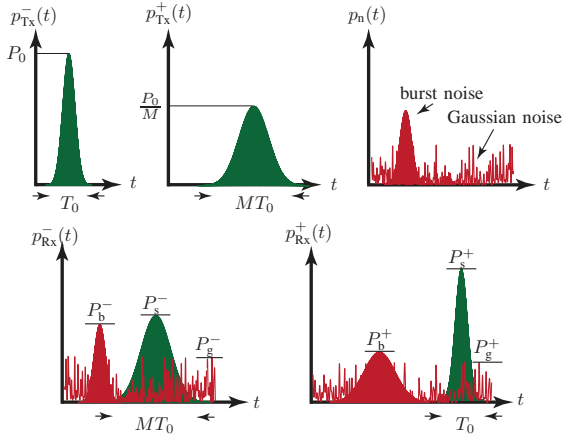}
{\psfrag{t}[c][c][0.7]{$t$}
\psfrag{o}[c][c][0.7]{$p^+_\text{Rx}(t)$}
\psfrag{y}[c][c][0.7]{$p^-_\text{Rx}(t)$}
\psfrag{n}[c][c][0.7]{$p_\text{n}(t)$}
\psfrag{e}[c][c][0.7]{$p^-_\text{Tx}(t)$}
\psfrag{f}[c][c][0.7]{$p^+_\text{Tx}(t)$}
\psfrag{r}[c][c][0.7]{SNR}
\psfrag{q}[c][c][0.7]{$P_0$}
\psfrag{g}[c][c][0.8]{$\frac{P_0}{M}$}
\psfrag{l}[c][c][0.7]{burst noise}
\psfrag{m}[c][c][0.7]{Gaussian noise}
\psfrag{r}[c][c][0.7]{$P^-_\text{b}$}
\psfrag{v}[c][c][0.7]{$P^+_\text{b}$}
\psfrag{s}[c][c][0.7]{$P^-_\text{s}$}
\psfrag{u}[c][c][0.7]{$P^+_\text{s}$}
\psfrag{w}[c][c][0.7]{$P^-_\text{g}$}
\psfrag{x}[c][c][0.7]{$P^+_\text{g}$}
\psfrag{0}[c][c][0.7]{$T_0$}
\psfrag{1}[c][c][0.7]{$MT_0$}}
\caption{Powers of the signals in Fig.~\ref{Fig:block}.} \label{Fig:principle}
\end{figure}

Let us now examine in more details how the signal is enhanced compared to noise and subsequently determine the SNR of the system. Figure~\ref{Fig:principle} shows the powers of the signals in Fig.~\ref{Fig:block} that are relevant to this end.

First consider the scenario of a noise consisting only of burst noise, with the same carrier frequency and the same bandwidth as the signal. The signal to noise ratio at the input of the down-chirp phaser in the receiver is then
\begin{equation}
   \text{SNR}^-_\text{b}=\frac{P^-_\text{s}}{P^-_\text{b}},
\label{Eq:sbnri}
\end{equation}
where $P^-_\text{s}$ is peak power of the up-chirped signal and  $P^-_\text{b}$ is the peak power of the burst noise. As the signal passes through the phaser, its peak power reduces by a factor $M$ due to time spreading and, consequently, the power at the output of the phaser is $P^+_\text{b}=P^-_\text{b}/M$. In contrast, the signal is enhanced by the factor $M$, due to dispersion compensation, so that the peak power at the output of the phaser is $P^+_\text{s}=MP^-_\text{s}$. As result, the output signal to burst noise ratio is
\begin{equation}
   \text{SNR}^+_\text{b}
   =\frac{P^+_\text{s}}{P^+_\text{b}}
   =\frac{MP^-_\text{s}}{P^-_\text{b}/M}
   =M^2~\frac{P^-_\text{s}}{P^-_\text{b}}=M^2\text{SNR}^-_\text{b},
\label{Eq:sbnro2}
\end{equation}
revealing that the burst noise has been enhanced by a factor~$M^2$.

Consider now a second scenario, where the noise is a Gaussian noise only. The signal to noise ratio at the input of the down-chirp phaser in the receiver is then
\begin{equation}
   \text{SNR}^-_\text{g}=\frac{P^-_\text{s}}{P^-_\text{g}},
\label{Eq:sbnrio}
\end{equation}
where $P^-_\text{s}$ is again the peak power of the up-chirped signal and $P^-_\text{g}$ is the peak power of the Gaussian noise. As in the previous scenario, the signal power is enhanced to $P^+_\text{s}=MP^-_\text{s}$. However, the power of the Gaussian noise does not change, i.e. $P^+_\text{g}=P^-_\text{g}$, since the effect of the phaser on such a noise is only to change its phase randomness. As a result, the signal to gaussian noise ratio at the output of the phaser is
\begin{equation}
   \text{SNR}^+_\text{g}
   =\frac{P^+_\text{s}}{P^+_\text{g}}
   =M\frac{P^-_\text{s}}{P^-_\text{g}}
   =M\text{SNR}^-_\text{g},
\label{Eq:snro}
\end{equation}
indicating that the signal to gaussian noise has been enhanced by a factor $M$.

\section{Experimental Demonstration}\label{sec:exp_dem}

\begin{figure}[h!]
\begin{center}
\subfigure[]{\label{Fig:up-chirp}
\centering
\psfragfig*[width=0.9\columnwidth]{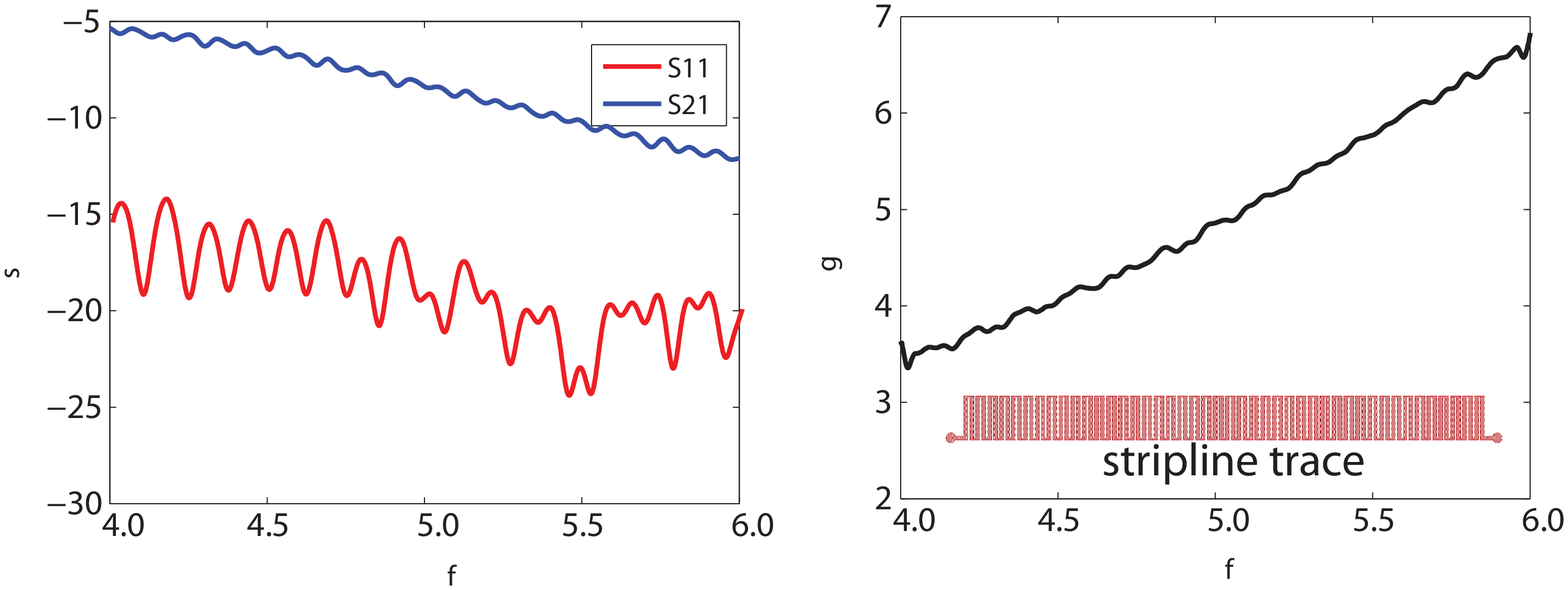}
{\psfrag{f}[c][c][0.7]{Frequency (GHz)}
\psfrag{s}[c][c][0.7]{S-parameters(dB)}
\psfrag{g}[c][c][0.7]{Group delay (ns)}}}
\subfigure[]{\label{Fig:down-chirp}
\centering
\psfragfig*[width=0.9\columnwidth]{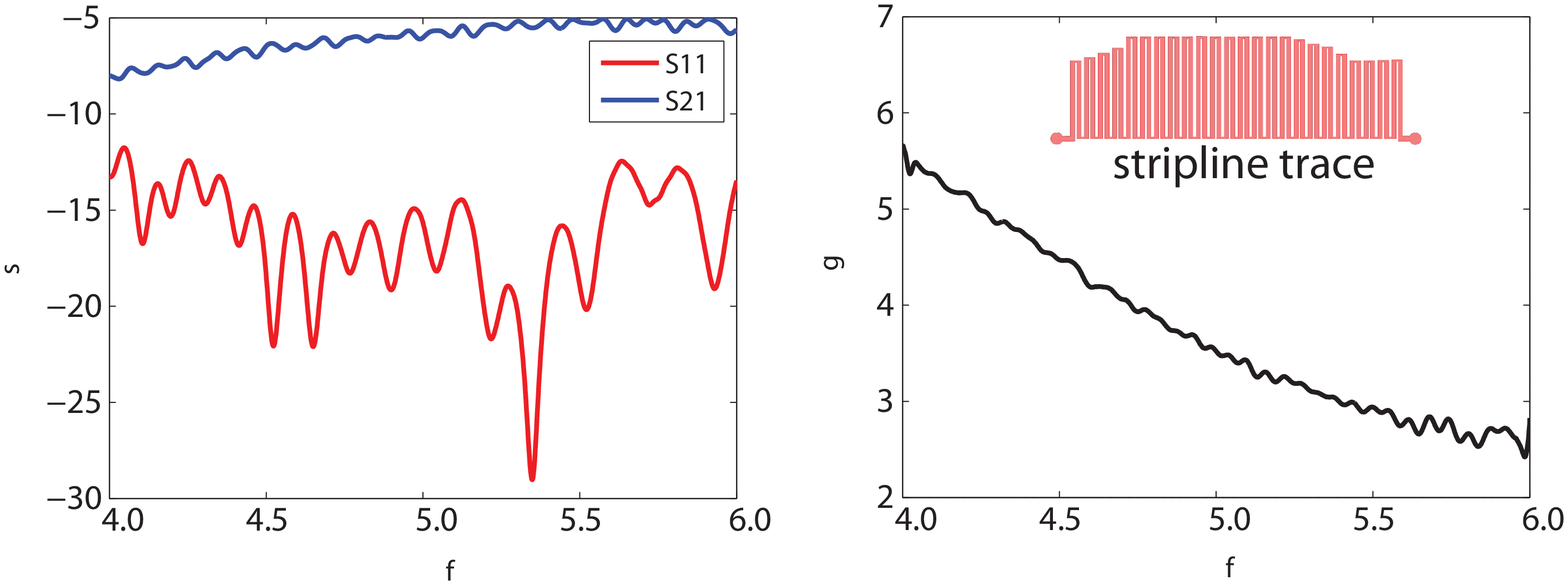}
{\psfrag{f}[c][c][0.7]{Frequency (GHz)}
\psfrag{s}[c][c][0.7]{S-parameters(dB)}
\psfrag{g}[c][c][0.7]{Group delay (ns)}}}
\caption{Measured responses for the two linear group-delay phasers implemented in cascaded C-section stripline technology for the experimental demonstration~\cite{Gupta_TMTT_09_2010}. The substrates are Rogers~3602 substrates, with $\epsilon_r=6.15$ and $\tan\delta = 0.0027$. The insets show the traces of the stripline structures. (a)~Up-chirp phaser, slope~=$+1.602$~ns/GHz ($=0.255$ns$^2/$rad). (b)~Down-chirp phaser, slope~$=-1.602$~ns/GHz.}\label{Fig:phaser}
\end{center}
\end{figure}

The experimental demonstration in this paper will resort to cascaded C-section phasers implemented in stripline technology following the synthesis procedures presented in~\cite{Gupta_TMTT_09_2010}. The traces of the designed phasers, and their measured S-parameter and group delay responses are plotted in Fig.~\ref{Fig:phaser}. As required, the two phasers were designed so as to provide opposite-slope and mutually compensating linear chirp responses over the frequency range of the signals. The stretching factor of the phasers is $M=(T_0+s\Delta\omega)/T_0=$2.38.

\begin{figure}[h!]
\centering
\psfragfig*[width=.98\columnwidth]{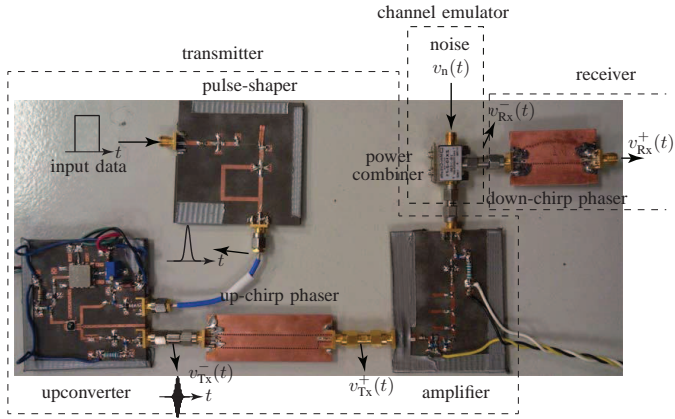}
{\psfrag{o}[c][c][0.7]{$v^+_\text{Rx}(t)$}
\psfrag{y}[c][c][0.7]{$v^-_\text{Rx}(t)$}
\psfrag{n}[c][c][0.7]{$v_\text{n}(t)$}
\psfrag{d}[c][c][0.7]{$\text{down-chirp phaser}$}
\psfrag{q}[c][c][0.7]{$\text{transmitter}$}
\psfrag{r}[c][c][0.7]{$\text{receiver}$}
\psfrag{p}[c][c][0.7]{$\text{power}$}
\psfrag{x}[c][c][0.7]{$\text{combiner}$}
\psfrag{a}[c][c][0.7]{$\text{amplifier}$}
\psfrag{j}[c][c][0.7]{$\text{pulse-shaper}$}
\psfrag{h}[c][c][0.7]{$\text{input data}$}
\psfrag{z}[c][c][0.7]{$\text{noise}$}
\psfrag{c}[c][c][0.7]{$\text{upconverter}$}
\psfrag{e}[c][c][0.7]{$v^-_\text{Tx}(t)$}
\psfrag{f}[c][c][0.7]{$v^+_\text{Tx}(t)$}
\psfrag{u}[c][c][0.7]{up-chirp phaser}
\psfrag{t}[c][c][0.7]{$t$}
\psfrag{g}[c][c][0.7]{channel emulator}}
\caption{Experimental transceiver system using the phasers in Fig.~\ref{Fig:phaser}.} \label{Fig:proto}
\end{figure}

The complete experimental transceiver system is shown in Fig.~\ref{Fig:proto}. For the sake of the proof-of-concept, the noisy channel is emulated by a power combiner that injects into the receiver the sum of the signal produced by the transmitter and of the noise. The input test signal, $g(t)$ [see Fig.~\ref{Fig:tx}] is a quasi-Gaussian pulse, of duration $T_0=1.5$~ns and bandwidth $\Delta f=$1.3~GHz, produced by the pulse-shaper from a rectangular input data pulse. The upconverter modulates the Gaussian pulse at 5~GHz, producing $v_\text{Tx}^-(t)$. The phaser up-chirps this signal into $v_\text{Tx}^+(t)$. Hence, given $\Delta\omega=2\pi\times1.3=$8.16~rad/ns and $s=0.255~$ns$^2/$rad, the total duration of $v_\text{Tx}^+(t)$ is $T=T_0+\Delta T=3.6$~ns ($\Delta T=s\Delta\omega=2.1$~ns). Finally, this signal is boosted by an amplifier and its magnitude [see Fig.~\ref{Fig:up-chirp}] is equalized~\cite{Equalizer} to avoid signal distortion~\cite{ASP-Magazine}. The resulted transmit signal is then passed through the channel emulator (power combiner with noise addition). The receiver essentially consists in the down-chirp phaser, which transforms the received noisy signal $v_\text{Rx}^-(t)+v_n(t)$ into the demodulated signal $v_\text{Rx}^+(t)$, that is visualized on a real-time oscilloscope.

\begin{figure}[h!]
\begin{center}
\subfigure[]{\label{Fig:burst-in}
\centering
\psfragfig*[width=0.47\columnwidth]{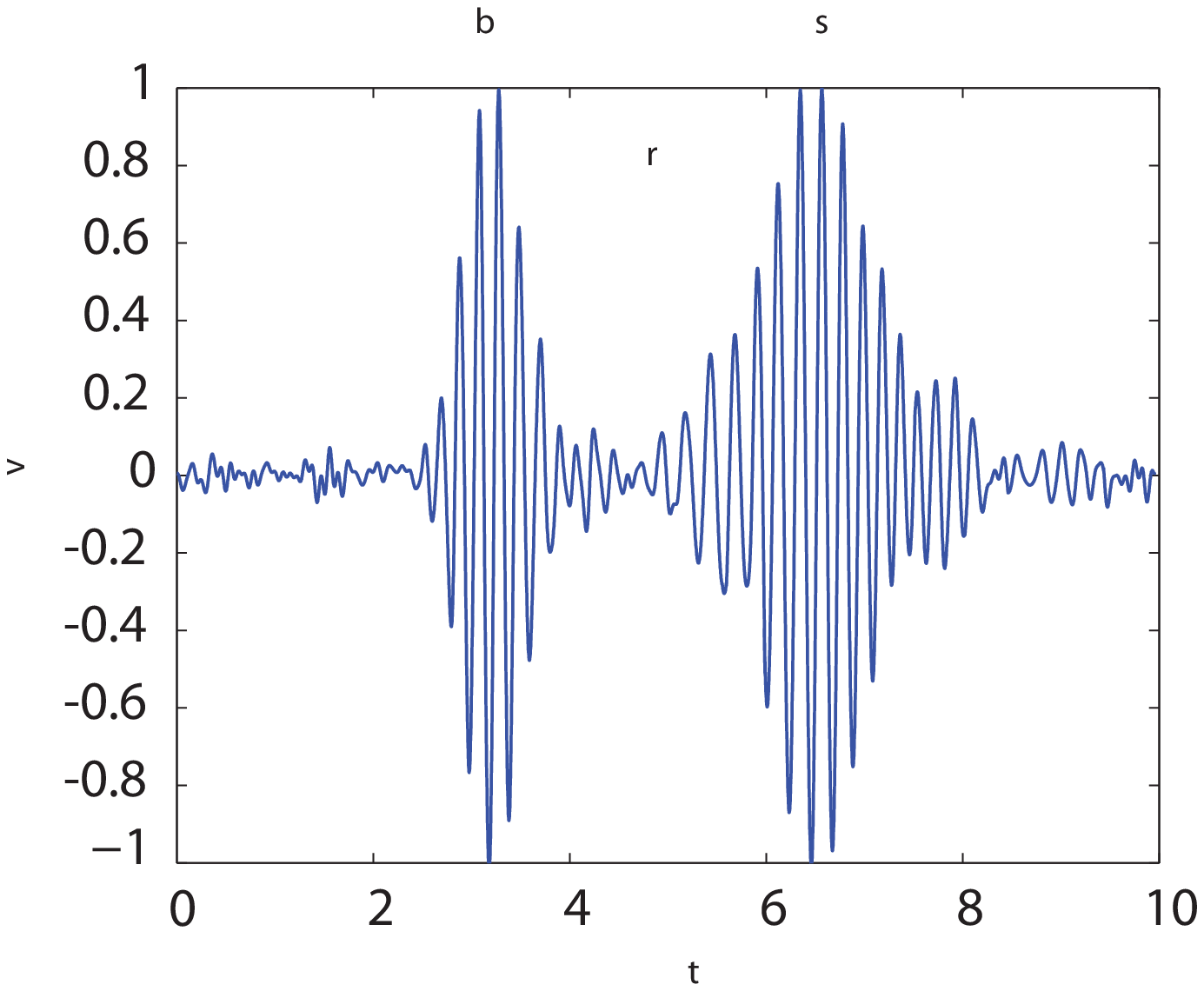}
{\psfrag{t}[c][c][.7]{Time (ns)}
\psfrag{v}[c][c][.7]{Normalized voltage}
\psfrag{b}[c][c][0.7]{burst noise}
\psfrag{r}[c][c][0.6]{$\text{SNR}^-_b=$0~dB}
\psfrag{s}[c][c][0.7]{signal}}}
\subfigure[]{\label{Fig:burst-out}
\centering
\psfragfig*[width=0.47\columnwidth]{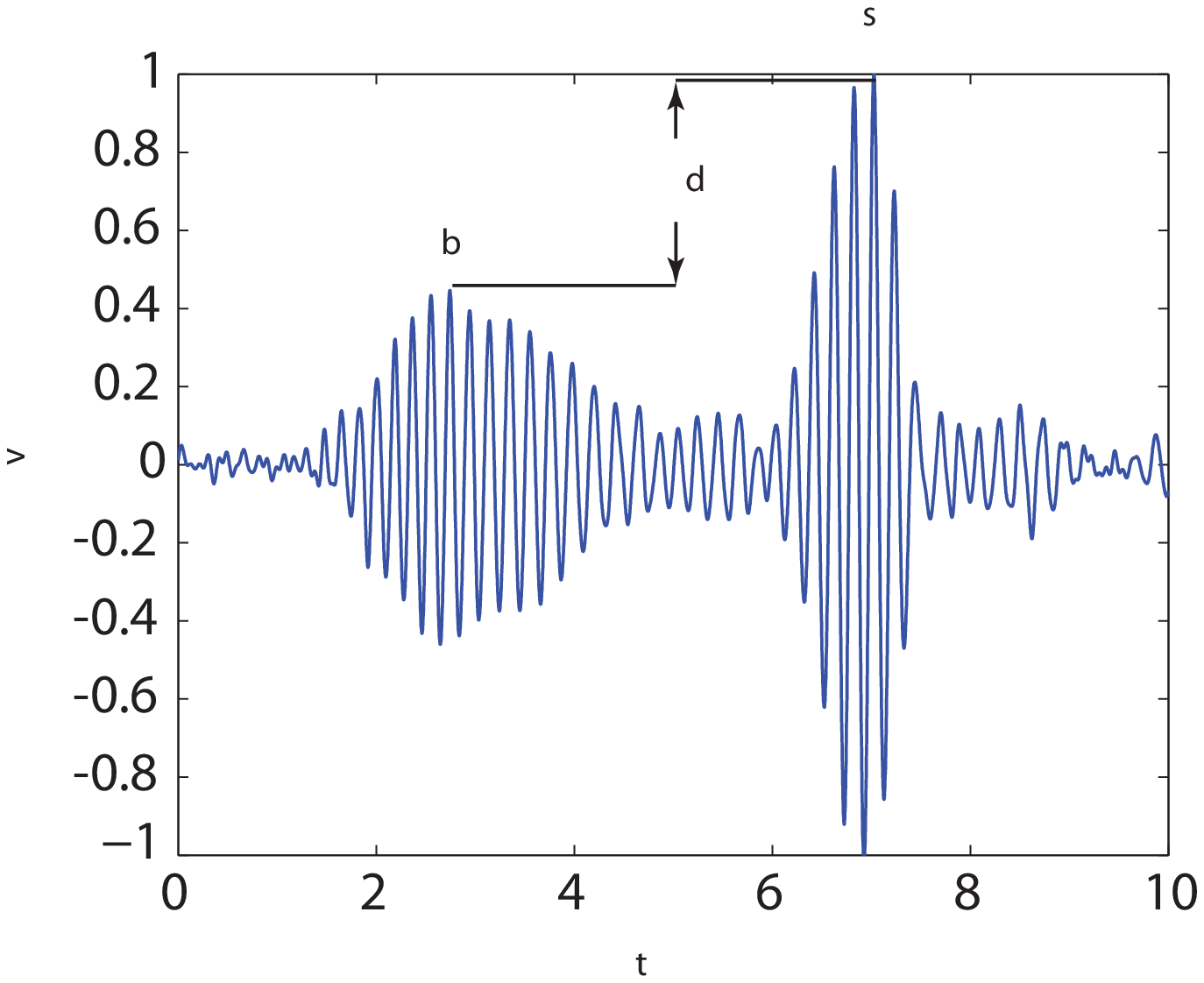}
{\psfrag{t}[c][c][.7]{Time (ns)}
\psfrag{v}[c][c][.7]{Normalized voltage}
\psfrag{b}[c][c][0.7]{burst noise}
\psfrag{s}[c][c][0.7]{signal}
\psfrag{d}[c][c][0.6]{$\text{SNR}^+_b=$7~dB}}}
\caption{Experimental results of the received signal in presence of the burst noise. (a)~Before the down-chirp phaser, $v^-_\text{Rx}(t)$, representing $\text{SNR}^-_b=$0~dB. (b)~After the down-chirp phaser, $v^+_\text{Rx}(t)$, representing $\text{SNR}^+_b=$7~dB.}\label{Fig:burst-noise}
\end{center}
\end{figure}

Figure~\ref{Fig:burst-noise} shows the measured results for the received signal in the presence of burst noise. The burst noise magnitude is set such that its peak level equals that of the signal are equal at the input of the receiver ($\text{SNR}^-_b=$0~dB), as shown in Fig.~\ref{Fig:burst-in}. The signal to burst noise ratio at the output of the down-chirp phaser has been enhanced by 7~dB, as shown in Fig.~\ref{Fig:burst-out}. The theoretical result for the SNR enhancement in~\eqref{Eq:sbnro2} is found to be $M^2=$5.66 ($=$7.52~dB), which is in close agreement with the measured SNR enhancement.
%
%
%

\begin{figure}[h!]
\begin{center}
\subfigure[]{\label{Fig:g-in}
\centering
\psfragfig*[width=0.48\columnwidth]{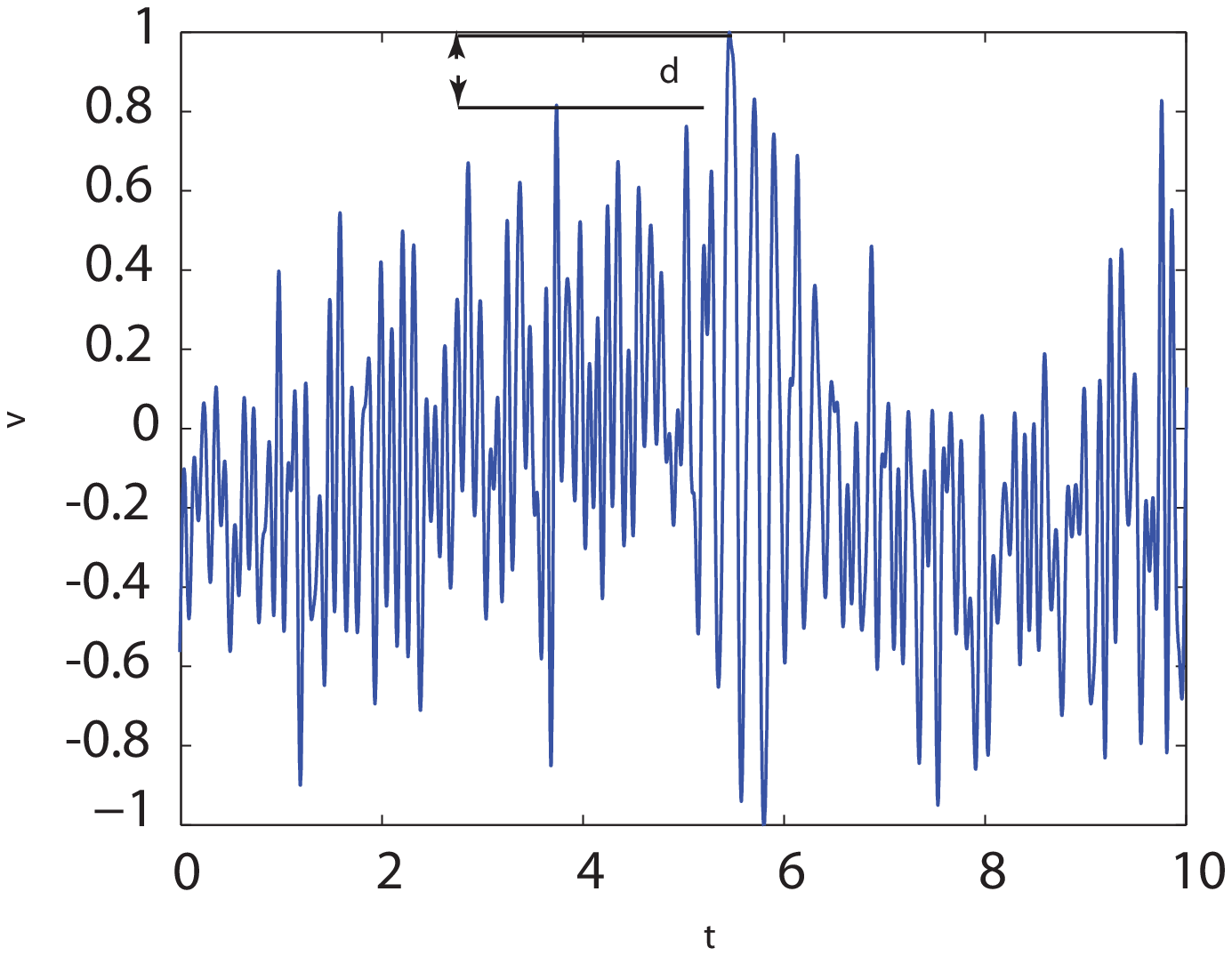}
{\psfrag{t}[c][c][.7]{Time (ns)}
\psfrag{v}[c][c][.7]{Normalized voltage}
\psfrag{b}[c][c][0.7]{burst noise}
\psfrag{e}[c][c][0.6]{overlapped noise}
\psfrag{f}[c][c][0.6]{and signal}
\psfrag{d}[c][c][0.6]{$\text{SNR}^-_g=$1.9~dB}
\psfrag{s}[c][c][0.7]{signal}}}
\subfigure[]{\label{Fig:g-out}
\centering
\psfragfig*[width=0.48\columnwidth]{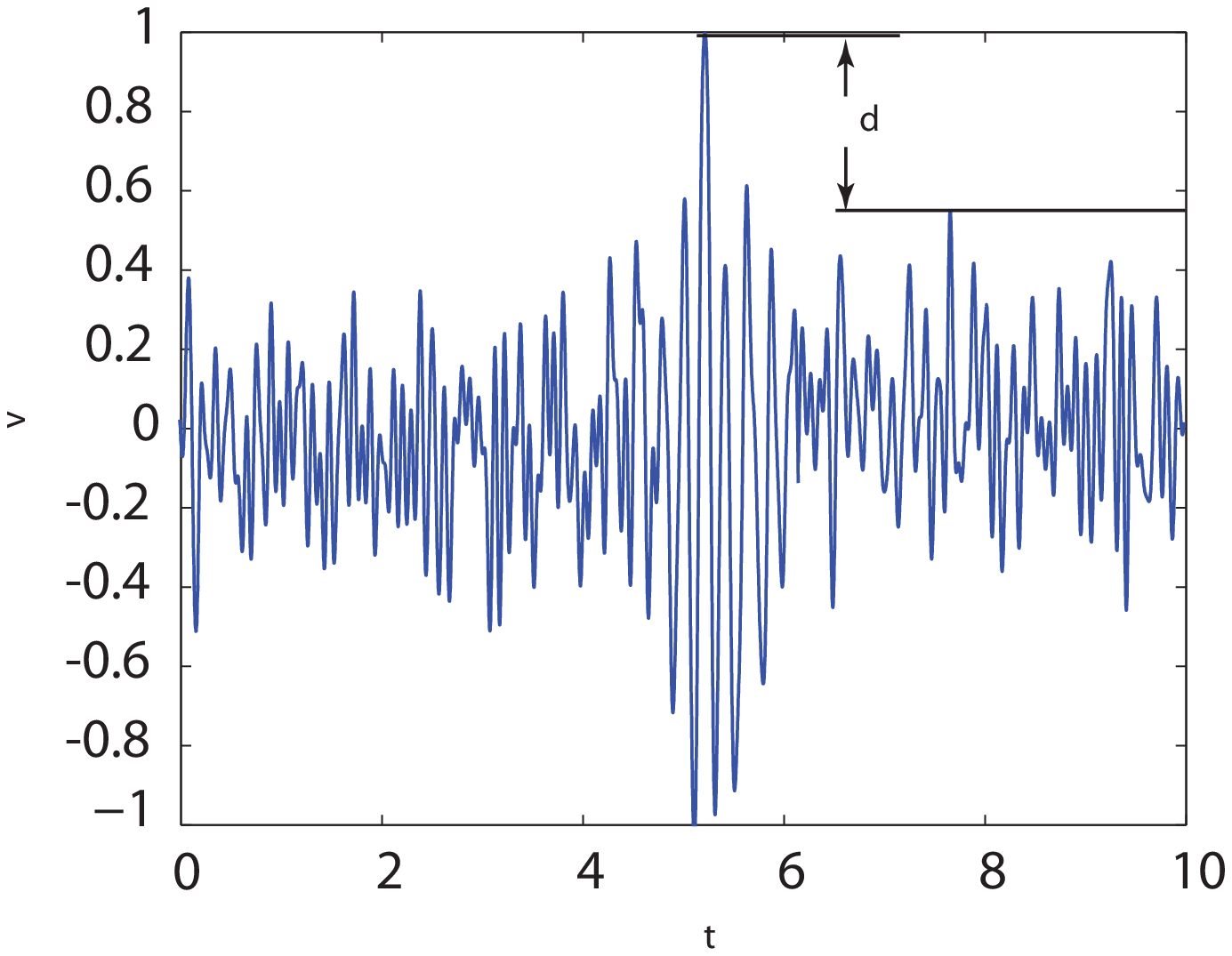}
{\psfrag{t}[c][c][.7]{Time (ns)}
\psfrag{v}[c][c][.7]{Normalized voltage}
\psfrag{b}[c][c][0.7]{burst noise}
\psfrag{s}[c][c][0.7]{signal}
\psfrag{d}[c][c][0.6]{$\text{SNR}^+_g=$5.2~dB}}}
\caption{Experimental results of the received signal in presence of the Gaussian noise. (a)~Before the down-chirp phaser, $v^-_\text{Rx}(t)$, representing $\text{SNR}^-_g=$1.9~dB. (b)~After the down-chirp phaser, $v^+_\text{Rx}(t)$, representing $\text{SNR}^+_g=$5.2~dB.}\label{Fig:Gaussian-noise}
\end{center}
\end{figure}

The measured results for the received signal to Gaussian noise is shown in Fig.~\ref{Fig:Gaussian-noise}. Here, the signal to Gaussian noise ratio is set to $\text{SNR}^-_g=$1.9~dB, as shown Fig.~\ref{Fig:g-in}. As shown in Fig.~\ref{Fig:g-out}, the signal to Gaussian noise ratio at the output of the down-chirp phaser is enhanced by 3.3~dB. This is again in close agreement with the theoretical prediction of~\eqref{Eq:snro}, $M=$2.38 (=3.76~dB).
%
%
%
%
\section{Conclusions}
We have proposed and experimentally demonstrated the concept of SNR enhancement in impulse radio transceivers using phasers of opposite chirping slopes. It has been shown that SNR enhancement by factors $M^2$ and $M$ are achieved for burst noise and Gaussian noise, respectively, where $M$ is the stretching factor of the phasers. The transceiver system is simple, low-cost and frequency scalable, and may therefore be suitable for broadband impulse radio ranging and communication applications.
\normalem
\bibliographystyle{IEEEtran}
\bibliography{ReferenceList}
\end{document}